\documentclass{aa501}

\usepackage{graphics}

\def\fh{\hbox{$.\!\!^{\rm h}$}}

\def \hi {H\,{\sc i~}} 
\def \hii {H\,{\sc ii~}} 

\def\kms{km\,s$^{-1}$} 
 
\def\mjyb{mJy~beam$^{-1}$}
\def\mas{mag~arcsec$^{-2}$}

\begin{document} 
\title{Status of \hi searches for CHVCs beyond the Local Group} 
\titlerunning{Searches for CHVCs beyond the Local Group} 
\author{R. Braun\inst{1} and W. B. Burton\inst{2} } 
\institute{Netherlands Foundation for Research in Astronomy, P.O. Box 2, 
7990 AA Dwingeloo, The Netherlands 
\and Sterrewacht Leiden, P.O. Box 9513, 2300 RA Leiden, The Netherlands } 

\date{Received mmddyy/ accepted mmddyy} 
\offprints{R. Braun} 

\abstract{
  Growing evidence supports the suggestion that the compact
  high--velocity clouds of \hi (CHVCs) are located throughout the Local
  Group and continue to fuel galactic evolution.  Recent distance
  estimates to individual objects lie in the range 150--850~kpc,
  implying an \hi mass range of $10^{5.7}\rightarrow10^{7.2}$ M$_\odot$
  together with sizes ranging from 2--12~kpc, while the average
  linewidth is 30~\kms\ FWHM. It is natural to ask whether objects of
  this type would not already have been seen by current blind \hi
  surveys of the extragalactic sky. We consider the properties of the
  deepest published surveys of this type and conclude that the
  sensitivity and coverage have now begun to reveal the high mass end
  of the distribution ($M_{\rm HI}>10^{7}$ M$_\odot$). Achieving
  detection limits an order of magnitude deeper should finally enable
  direct study of these systems beyond the Local Group, or definitively
  rule out their existence.  
}
\maketitle 
 
\section{Introduction} 
\label{intro}

Many aspects of the high--velocity cloud \hi emission detected in
large--area surveys at radial velocites between about $-$450 and +400
\kms~can be understood if this emission is associated with a population
of low--mass, dark--matter--dominated structures distributed throughout
the Local Group. This scenario was rejuvenated in a modern context by
Blitz et al. (\cite{blit99}), based on the HVC tabulation of Wakker \&
van Woerden (\cite{wakk91c}) and data of the Leiden/Dwingeloo \hi
Survey (LDS) of Hartmann \& Burton (\cite{hart97}), and by Braun \&
Burton (\cite{brau99}), based on their extraction of compact, isolated
HVCs from the LDS and confirmed with independent data.  In the Local Group
scenario, extended HVC complexes are nearby objects currently merging
(or soon to merge) with the Galaxy, while the compact isolated high--velocity
clouds (CHVCs) are the distant counterparts of the complexes which have
as yet not strongly interacted with the massive Local Group
constituents.

At least one of the HVC complexes (Chain A) lies at a distance between
about 4 and 10~kpc (van Woerden et al. \cite{vanw99}); another (Complex
C) has a metallicity of only 0.09 solar (Wakker et al. \cite{wakk99}),
implying an extragalactic origin even if the complex is currently
nearby.  If located at $D\sim$15~kpc, the \hi mass of Complex~C is
$10^7$~M$_\odot$ and its linear size is 15~kpc.  Further evidence for
the extragalactic nature of the CHVCs follows from the metallicity
measurement of 0.05 solar for CHVC\,125+41$-$207 by Braun \& Burton
(\cite{brau00}). The derived distance of this same object, which
follows from a comparison of the \hi column and volume densities, is
650$\pm$300~kpc. Comparably large distances and high dark--to--visible
mass ratios (10 to 50) are also implied by the rotation signatures
displayed by many CHVCs (Braun \& Burton \cite{brau00}).

In the direct vicinity of massive galaxies we expect that these objects
will live but shortly before tidal disruption and merger ensue. Such
ongoing mergers are seen regularly in nearby galaxies, e.g. LMC (Luks
\& Rohlfs \cite{luks92}), M33 (Deul \& van der Hulst \cite{deul87}),
M101 (Kamphuis \cite{kamp93}), NGC\,628 (Kamphuis \& Briggs
\cite{kamp92}). Gas masses of a few times 10$^6$ to more than
10$^8$~M$_\odot$ and physical dimensions of 10 to 20~kpc are observed.

Counterparts to the CHVCs beyond the Local Group are more difficult to
constrain.  We argue here that current surveys are only beginning to
achieve the appropriate combination of sensitivity and spatial
coverage. 

\section{CHVC populations, distance, mass and size} 
\label{sec:pops}

Where and in what numbers might CHVC populations be expected to occur?
Low--mass objects which have so far escaped assimilation by high--mass
galaxies might be found in the vicinity of substantial,
dynamically--young mass concentrations at distances less than the
turn--around radius of the local over--density. In the case of the
Local Group, this radius is at about 1.2~Mpc (Courteau \& van den Bergh
\cite{cour99}) for a total mass of some 2.3$\times10^{12}$~M$_\odot$.
The turn--around radius scales as the cube root of mass (Sandage
\cite{sand86}). For field galaxies with a typical luminosity of a few
$10^9$~L$_\odot$ and a mass of a few 10$^{10}$~M$_\odot$ the
turn--around radius would lie at about 250~kpc.  The total number of
associated low--mass objects in a dynamically young system is expected
to scale linearly with total mass (Klypin et al.  \cite{klyp99}). 
Relative to the Local Group, we would then expect some 100 times fewer
objects associated with a single L$^*$ galaxy in the field.

What is the typical \hi mass, linewidth and linear size expected for a
population of CHVCs? The average integrated flux of the 65 objects in
the CHVC sample of Braun \& Burton (\cite{brau99}) is 100 with a
dispersion of 75~Jy\,\kms; the average linewidth 30~\kms\ FWHM; and the
average FWHM angular size is 50 with a dispersion of 25~arcmin when
imaged with a 36~arcmin beam.  Only a small number of distance
estimates for CHVCs are currently in hand and even these are indirect,
relying on an assumption about the typical thermal pressure of each
source.

\subsection{CHVC distances utilizing compact opaque cores}

For objects that have compact opaque cores (in which the \hi\
brightness temperature in emission is comparable to the gas kinetic
temperature) with a column density enhancement, $dN_{\rm HI}$, of
angular size, $\theta$, it is possible to estimate the distance from
$D_{\rm Core}~=~dN_{\rm HI}^{\rm Core}/(f n_{\rm HI} \theta)$, assuming
crude spherical symmetry, if we have an estimate of the \hi\ volume
density, $n_{\rm HI}$ and its volume filling factor along the
line-of-sight, $f$. Since the thermal pressure is given by
$P~=~k_Bn_{\rm HI}T_k$, we can rewrite the expression for distance as,
\begin{eqnarray}
D_{\rm Core} =  {670 \over f} \biggl({dN_{\rm HI}^{\rm Core} 
\over 10^{21}{\rm cm}^{-2}}\biggr) 
\biggl({T_k \over 100\ {\rm K}}\biggr) \ \ \ \ \ \ \ \ \ \ \ \  &  \nonumber \\
\biggl({P/k \over 100~{\rm cm}^{-3}{\rm K}}\biggr)^{-1}   
\biggl({\theta \over 100''}\biggr)^{-1} & {\rm kpc.} 
\label{eqn:Dcore}
\end{eqnarray}
In general, only a crude lower limit to the
distance follows from this approach since the filling factor is
difficult to determine and might be substantially less than
unity. However, in those cases where the core \hi\ is essentially
optically thick, a filling factor near unity can be plausibly assumed. 

This method can perhaps best be demonstrated by using M31 as an
example. The excess column density, $dN_{\rm HI}^{\rm Core} =
5\pm1\times10^{21}$cm$^{-2}$, and angular size, $\theta =
60\pm20$~arcsec, of opaque \hi\ clumps in the North-East half of M31
can be estimated from Fig.~5 of Braun \& Walterbos (\cite{brau92}). The
\hi\ kinetic temperature, $T_k = 175\pm25$~K, and estimated thermal
pressure, $P/k = 1500\pm500$~cm$^{-3}$K, of the M31 mid-disk are
tabulated in Table~4 of the same reference. This yields the distance
estimate: $D_{M31} = 650\pm220$~kpc, assuming a volume filling factor
for these clumps of unity. While rather crude, this approach gives a
plausible value for the distance to M31.

Opaque \hi\ cores have only been detected in one CHVC to date, namely
CHVC125+41$-$207 (Braun \& Burton \cite{brau00}), where an excess
column density, $dN_{\rm HI}^{\rm Core} = 1.0\pm0.2\times10^{21}$cm$^{-2}$,
and angular size, $\theta = 90\pm15$~arcsec, is measured in the \hi\
clumps. A good estimate of the kinetic temperature follows from the
measured linewidth in these clumps of only 2.0~\kms\ FWHM, for which
$T_k = 85\pm10$~K, while a peak brightness temperature of 75~K is
observed, suggesting that a filling factor near unity might be
considered.

What must still be estimated to employ eqn.\ref{eqn:Dcore} is the
appropriate thermal pressure.  Although the pressure, $P/k$, in the
solar neighborhood of the Galaxy is well-determined at about
2000~cm$^{-3}$K, it is expected to decline rapidly with distance from
the Galactic plane (eg. Wolfire et al. \cite{wolf95}) such that beyond
about 10~kpc we should encounter values, $P/k = 100$~cm$^{-3}$K or less
due to the Galactic halo. 

Within the inter-galactic medium (IGM) of the Local Group, it is
difficult to determine what values of thermal pressure might be
encountered. Current instrumentation has allowed diffuse X-ray emission
to be detected from the IGM in several ``poor'' galaxy groups
(eg. Mulchaey et al. \cite{mulc96}). It must be noted that even the
poorest of the currently detected groups is substantially richer than
the Local Group and all have proven to be lacking in spirals. The X-ray
detected systems have derived gas masses of about 10$^{13}$M$_\odot$
(comparable to or less than the mass in galaxies), temperatures of
about 1~keV (corresponding to 10$^7$~K) and radii of about
500~kpc. Mulchaey et al. suggest that spiral-dominated galaxy groups
are likely to have lower X-ray temperatures, of perhaps
0.2~keV. Assuming a correspondingly lower total gas mass of about
10$^{12}$M$_\odot$ and radii between 500 and 1000~kpc such as might
apply to the Local Group, would imply a Local Group IGM thermal
pressures in the range 160 to 20~cm$^{-3}$K.

Braun \& Burton (\cite{brau00}) have considered the total pressure
implied by assuming hydrostatic equilibrium of a self-gravitating
gaseous disk. The self-gravity of the WNM in a diffuse disk system
accounts for a contribution, $P/k = 40(N_{\rm HI}^{\rm
Halo}(0)/10^{20}$cm$^{-2}$)$^2$~cm$^{-3}$K in the absence of a
significant dark matter contribution to the disk mass. The peak column
densities of WNM found in CHVCs are in the range
10$^{19.5}$--10$^{20}$cm$^{-2}$ (Burton et al \cite{burt00}).

The resulting thermal pressure, the sum of self-gravitating and
external components, at the interface of the warm neutral
halo with the cool condensed cores, of about $P/k = 100\pm50$~cm$^{-3}$K
is in very good agreement with that expected theoretically (Wolfire et
al. \cite{wolf95}. This transition pressure is found to apply over a
wide range of physical conditions (metal abundance, radiation field and
dust content). Taking the value, $P/k = 100\pm50$~cm$^{-3}$K,
allows calculation of the distance to CHVC125+41$-$207, yielding
$D_{\rm C125} = 600\pm300$~kpc. 

\subsection{CHVC distances utilizing edge profiles}

A second method of distance estimation (Burton et al. \cite{burt00})
makes use of the column density profiles of the diffuse WNM halos of
these objects. These halos can be well-described by the sky-plane
projection of a spherical exponential distribution of the \hi volume
density,
\begin{equation}
n_{\rm HI}(r) = n_{\rm o}e^{-r/h}
\label{eqn:nofr}
\end{equation}
in terms of the radial distance, $r$, and exponential scale length,
$h$. The projected column density distribution is then,
\begin{equation}
N_{\rm HI}^{\rm Halo}(r) = 2 h n_{\rm o} 
\biggl[ {r \over h}K_1 \biggl( { r \over h}\biggr)
\biggr]
\label{eqn:Nofr}
\end{equation}
where $K_1$ is the modified Bessel function of order 1. In this
formulation, the peak halo column density is given simply by,
\begin{equation}
N_{\rm HI}^{Halo}(0) = 2 h n_{\rm o} 
\label{eqn:N0}
\end{equation}
The distance of the source can then be calculated from,
\begin{equation}
D_{\rm Halo} = {N_{\rm HI}^{\rm Halo}(0) \over 2 h n_o} = 
{N_{\rm HI}^{\rm Halo}(0) k_B T_k^{\rm Halo} \over 2 h P} 
\label{eqn:DiH}
\end{equation}
where the implicit assumption has been made of a unit filling factor
for the halo gas. If we assign the measured temperature of
$T_k^{\rm Halo}=10^4$~K to the halo gas this becomes,
\begin{eqnarray}
D_{\rm Halo} =  335 \biggl({N_{\rm HI}^{\rm Halo}(0) 
\over 10^{19}{\rm cm}^{-2}}\biggr) 
\ \ \ \ \ \ \ \ \ \ \ \ \ \ \ \ \ \ \nonumber & \\
\biggl({P/k \over 100~{\rm cm}^{-3}{\rm K}}\biggr)^{-1} 
\biggl({h \over100''}\biggr)^{-1} &  {\rm kpc}.
\label{eqn:DisH}
\end{eqnarray}
The peak halo column densities, $N_{\rm HI}^{\rm Halo}(0)$, and
scale-lengths, $h$, can be determined from sensitive total power
observations, like the deep Arecibo cross-cuts obtained for a sample of
ten objects by Burton et al. (\cite{burt00}). Assuming, as before, a
nominal thermal pressure of $P/k = 100\pm50$~cm$^{-3}$K at the
interface of the cool and warm neutral \hi, yields distance estimates
which range from 150 to 850~kpc for the various objects studied.

An independent method of distance estimation (Burton et
al. \cite{burt00}) follows from simply equating the angular \hi
scale-lengths observed in the CHVCs with the mean physcial
scale-length, $h = 1.1\pm0.4$ kpc, measured in a sample of low mass
dwarf galaxies. This yields distances in the range 320 to 730~kpc.

Apparently, a wide range of distances, spanning almost an order of
magnitude, is indicated for the Local Group CHVC population. If a
distance range of 150$\rightarrow$850~kpc is representative for the
CHVCs, then the corresponding \hi mass range is $M_{\rm
HI}=10^{5.7}\rightarrow10^{7.2}$ M$_\odot$. The distribution of masses
over this range is not yet known, but is likely to be more heavily
populated at the low end than the high.  The critical \hi mass
detection threshold is therefore about $10^{6}$~M$_\odot$, if a large
fraction of the population is to be detected.  If the typical radial
profile of the CHVCs were Gaussian, then the corresponding range of
FWHM linear sizes would be 2$\rightarrow$12~kpc. Since the edge
profiles are instead observed to be exponential, these objects will
appear slightly resolved when observed with any telescope beam broader
than that of the high column density cores of about 5--10~arcmin (Braun
\& Burton \cite{brau00}), corresponding to 1--2~kpc. Unfortunately,
with an exponential edge profile, the observed object ``size'' and
``total mass'' will depend not only on the resolution used, but also on
the sensitivity achieved. For a 2-D circular exponential distribution,
the fractional flux contained within a radius of one, two and three
scale-lengths is 26, 59 and 80\%. The implication is that only for beam
sizes greater than about 20~kpc, is it likely that the entire flux of
an object will be detected in cases of modest signal-to-noise ratio.

\section{Synthesis Surveys } 
\label{sec:synsur}

One method of achieving ``blind'' survey coverage for low mass objects
is to utilize the relatively large field-of-view of an interferometeric
array when observing the \hi distribution in nearby galaxies to obtain
serendipitous detection of nearby uncataloged systems.  Some aspects of
this problem have been considered previously by Blitz et
al. (\cite{blit99}). With a typical interferometric field of view of
30~arcmin, an area with linear size of 90~$D_{10}$ kpc (normalized to a
source at 10~Mpc) is probed with a single pointing. Since correlator
capacity is often a limiting factor for such observations, the typical
velocity coverage obtained is often only about 1.5 times the velocity
extent of the target galaxy.  Depending on the goals of a particular
observing program, this entire field and velocity coverage may not even
be analyzed; but if it were, the probability of intercepting an
associated object would be proportional to the total number of objects,
$N_{\rm Grp}$ for a galaxy group or $N_{\rm Gal}$ for an isolated
galaxy, as well as the fractional area and velocity extent
observed. Assuming that the velocity coverage is sufficient to sample
the entire associated population, this corresponds to
$N~=~0.002\,N_{\rm Grp}\,D_{10}^2$ for a distribution extending out to
1~Mpc (a distance comparable to the Local Group turn-around radius) or
$N~=~0.03\,N_{\rm Gal}\,D_{10}^2$ for an isolated galaxy. Only for
values of $N_{\rm Grp}$ as large as 500 or $N_{\rm Gal}$ = 30 would one
expect serendipitous detections in individual pointings that achieve a
detected mass sensitivity of better than about $10^6$~M$_\odot$ over
30~\kms\ with a beam size larger than about 20~kpc.

The required mass sensitivity for CHVC detection in external galaxy
groups has not yet been generally attained. Taylor et al.
(\cite{tayl95},\cite{tayl96}) have searched the fields of 21 apparently
isolated \hii galaxies and of 17 low surface brightness dwarf
galaxies. The targeted galaxies were at a typical distance of about
25~Mpc, where the half-power field-of-view is 225~kpc and the spatial
resolution about 8~kpc. The observations reached an average five sigma
mass sensitivity over 30~\kms\ of about $2.5\times10^7$ M$_\odot$ at the
field center in the case of the \hii galaxies and $1.7\times10^7$
M$_\odot$ in the case of the LSB dwarf galaxies. While a total of about
17 companion galaxies were discovered in this survey, all but one
(UM422C) were found to have optical counterparts detectable at surface
brightnesses of 23 \mas. Since the \hi detection thresholds have very
little overlap with the mass range of the Local Group CHVC population
this result can be readily understood.

Another sensitive search for \hi emission was carried out by Van Gorkom
et al. (\cite{vang93}) in the vicinity of Ly$\alpha$
clouds in a single $30'$ FWHM field centered on 3C\,273 and covering
$+840$ to $+1840$ \kms. The field is located some $10^\circ$ (or about
2.6~Mpc) from the center of the Virgo Cluster (at an assumed distance
of 15~Mpc). The rms noise in the center of their field corresponded to
a $5\sigma$ mass sensitivity of $2.5$--$4.5\times10^{6}$M$_\odot$, in a
single 41.6\kms~channel at linear resolutions of 1.5--4.2~kpc. Van
Gorkom et al. estimate a $5\sigma$ mass sensitivity at 10~kpc
resolution (to better match the expected CHVC size) of
$1.1\times10^{7}$M$_\odot$ at the field center. Spatially averaged over
the $30'$ FWHM Gaussian field-of-view yields a $5\sigma$ average mass
limit of $1.6\times10^{7}$M$_\odot$ over a region 0.013~Mpc$^2$ in
extent. The fractional area sampled relative to the impact parameter of
2.6~Mpc is 0.06\%.  No \hi emission was detected, implying an upper
limit of some 2000 uniformly distributed objects with $M_{\rm
  HI}>1.6\times10^7$ M$_\odot$, if the imaged field is representative.
We note that the Virgo Cluster is a dynamically evolved system, for
which the long term survival of a low mass population is uncertain.
But, in any case, the average mass sensitivity of this search is
insufficient to detect large numbers CHVC counterparts. 

More recently Pisano \& Wilcots (\cite{pisa99},\cite{pisa00}) have
surveyed the fields of 34 isolated galaxies with distances in the range
21 to 45~Mpc. With a linear resolution of 6--13~kpc, they achieve a
5$\sigma$ mass sensitivity over 30~\kms\ of $1.4\pm0.6\times10^7$
M$_\odot$ at each field center. This sensitivity begins to overlap with
the range of expected CHVC masses near the field center. However,
averaged over the primary beam out to the half power point,
corresponding to radii of 90--190~kpc, the mass sensitivity is
decreased to $2\pm1\times10^7$~M$_\odot$, where effectively no
detections are expected. Although several gas-rich companions are
detected in this survey, all but possibly one are accompanied by
optical counterparts.

\section{Total Power Surveys}
\label{sec:TPsur}

One of the earliest systematic searches for \hi in galaxy groups was
that of Lo \& Sargent (\cite{lo79}). Large regions of the sky in the
vicinity of three nearby galaxy groups were searched using the OVRO
40~m telescope for \hi emission features without obvious optical
counterparts. The 5$\sigma$ mass sensitivity for a 35~\kms\ FWHM
linewidth was $4\times10^7$~M$_\odot$ for a region of 132 square
degrees (or 0.4~Mpc$^2$) adjacent to M81, while mass limits of
$8\times10^7$ over 0.7~Mpc$^2$ and $46\times10^7$~M$_\odot$ over
5.7~Mpc$^2$ were obtained in 90 square degree regions near the CVnI
group and NGC~1023 respectively. A total of six gas-rich objects were
subsequently detected in follow-up work using the 100~m Effelsberg
telescope, although all of them do have (faint) optical
counterparts. Comparison with our mass estimates above indicates that
the non-detection of an extensive CHVC population is entirely in
keeping with the likely mass distribution of Local Group CHVCs.

\subsection{The Arecibo \hi Strip Survey} 
\label{sec:ahiss}

A sensitive, large area \hi survey has been the Arecibo \hi Strip
Survey (AHISS) by Zwaan et al.  (\cite{zwaa97}). In fact, Zwaan \&
Briggs (\cite{zwaa00}) claim that the AHISS would have made numerous
detections of CHVCs if they did populate the outer environments of
galaxies and galaxy groups. In order to critically assess this claim,
we have carefully considered the attributes of the survey which are
well-documented by Sorar (\cite{sora94}).

The spatial coverage of the AHISS consists of two
repeatedly--observed strips, one covering the right ascension interval
$18^{\rm h}$ to $05^{\rm h}$ at $\delta=14^\circ14'$, and the other
covering $01^{\rm h}$ to $10\fh7$ at $\delta = 23^\circ09'$. The
velocity coverage extends from $-700$ to +7500~\kms. The noise level of
the AHISS data depends on position as well as velocity for a number of
reasons. Firstly, varying integration times were obtained for different
positions. Secondly, the receiver gain varied significantly with
frequency. For the $\delta=14^\circ14'$ strip this implied a factor of
two degraded sensitivity at 0~\kms\ than at +5000~\kms, while for the
$\delta = 23^\circ09'$ strip, the degradation at 0~\kms\ was limited to
about 15\%.  And lastly, confusion and spectral baseline residuals due
to bright, extended emission from the Galaxy preclude detection of
objects within about $|V_{GSR}| < 200$~\kms.

The average rms level of the $\delta=14^\circ$ strip is 0.95~\mjyb\ at
16~\kms\ resolution at velocities near +5000~\kms, rising to 1.9~\mjyb\
at low recession velocites. For the $\delta=23^\circ$ strip, the
average noise level is 0.75~\mjyb\ at 16~\kms~resolution between about
+5000 and +1000~\kms, rising to 0.85~\mjyb\ at low recession velocities.

The spatial coverage of the AHISS is given by the track of the
telescope beam (about 180 arcsec FWHM at 1400~MHz) along the indicated
right ascension intervals. The maximum sensitivity, noted above, is
only obtained for completely unresolved sources which pass within the
central 20 arcsec or so of the beam response.  If instead we consider a
realistic search area comparable in width to the beam FWHM, then the
average detected flux is down by a factor of 0.81 for an approximately
Gaussian main lobe. This diminished response must be taken into account
when assessing the sensitivity for unresolved sources over the full 180
arcsec width. If instead, a source is significantly resolved by the
survey beam, then the fraction of detected flux at a given offset
position is given by the convolution of the beam and source shapes. In
the case of two circular Gaussians of similar FWHM, with random linear
offsets in the range $\pm$HWHM, the average detected flux is reduced by
a factor of 0.64 from the intrinsic flux.

The mass sensitivity of the AHISS averaged over the FWHM of the
telescope beam can now be assessed as a function of recession
velocity. For nearby systems, say at 5~Mpc distance, the telescope beam
corresponds to only 4.4~kpc and any detected system will be strongly
resolved. The rms noise averaged over the $\delta=14^\circ$ and
$23^\circ$ strips at V$_{GSR}$~=~300~\kms, is 1.4~\mjyb\ over 16~\kms,
yielding a 5$\sigma$ limit on {\it detected} signal strength of
160~\mjyb-\kms\ over the typical 32~\kms\ linewidth. The average {\it
intrinsic} signal strength that gives this response will be at least a
factor of 1/0.64 higher (due to offsets from the beam center) yielding
a 5$\sigma$ mass limit of $1.5\times10^6$~M$_\odot$. At this distance
the AHISS is indeed capable of detecting a large part of the expected
CHVC mass distribution, even with some further reduction in detected
flux due to the exponential edge profiles.

At a distance of 15~Mpc, the telescope beam coresponds to 13~kpc, which
is comparable to the expected source size. The rms noise averaged over
the $\delta=14^\circ$ and $23^\circ$ strips at V$_{GSR}$~=~1000~\kms,
is 1.1~\mjyb\ over 16~\kms, yielding a 5$\sigma$ limit on detected
signal strength of 125~\mjyb-\kms\ over the typical 32~\kms\ 
linewidth. The average intrinsic signal strength will again be higher
by the beam averaging factor of 1/0.64 (for the comparable beam and
source size case), yielding a 5$\sigma$ mass limit
of $1.0\times10^7$~M$_\odot$. This constitutes a practical distance
limit to the AHISS for expected detection of the upper mass end of the
CHVC mass distribution. 

We searched the Garcia (\cite{garc93}) catalog of galaxy groups for all
those in the range +200 to +1200~\kms~and within 1~Mpc projected
distance of the AHISS coverage. Only one cataloged galaxy group is
intercepted by the AHISS in this range, namely the NGC\,628 group
centered at $(\alpha,\delta,V)=(01^{\rm h}36^{\rm m},14^\circ55',+570$
\kms), at a distance of about 10~Mpc. There is no other instance of the
AHISS coverage passing near a group of galaxies within 15~Mpc.

The AHISS sensitivity in the NGC\,628 region is 1.4~\mjyb over 16~\kms,
while the telescope beam corresponds to about 9~kpc. With a beam
averaging factor of 1/0.64, this yields a 5$\sigma$ mass sensitivity of
$5.7\times10^6$~M$_\odot$ over the expected linewidth of 32~\kms.  This
sensitivity should be sufficient to allow detection of the more massive
members of a CHVC population, if they happened to lie within the survey
strip.  The AHISS coverage passes within 120~kpc of the nominal
NGC\,628 group center, so the group area intercepted by this strip
corresponds to a region about 9 kpc wide (for the 180 arcsec beam FWHM
at $D=10$~Mpc) and some 2~Mpc long. This area corresponds to some 0.6\%
of the total group area ($\pi R^2$) of about 3.1 Mpc$^2$. For a uniform
distribution of objects, the AHISS non--detection implies that the
total number of such objects exceeding $5.7\times10^6$~M$_\odot$ should
be less than about 300.

The next nearest group covered by the AHISS survey is the NGC\,3227
group at $(\alpha,\delta,V)=(10^{\rm h}19^{\rm m},21^\circ23',+1250$
\kms). Since this is covered in the $\delta=23^\circ$ strip at a
velocity greater than +1000~\kms, the flux
sensitivity is 0.75~\mjyb at 16~\kms~resolution.  The 5$\sigma$ mass
sensitivity at an estimated distance of 19.2~Mpc is then $M_{\rm
HI}=1.1\times10^7$ M$_\odot$ over 32~\kms. This sensitivity is
sufficient to detect only the most massive objects in a CHVC
population.  Since the survey coverage passes through the group at an
impact parameter of about 590~kpc, the 17~kpc beamwidth samples a
fractional group area of about 0.9\%. The AHISS non-detection implies
that the number of objects exceeding $1.1\times10^7$ M$_\odot$ should
be less than about 200.

We have also searched the LEDA database for all cataloged galaxies with
recession velocities (corrected for Virgo--centric infall) in the range
+200 to +1300~\kms~and projected separations of less than 250~kpc from
the AHISS strip coverage.  Apart from galaxies belonging to the
NGC\,628 group, and therefore already considered above, a total of 7
other galaxies lie this close to the AHISS coverage. Six of the seven
are low mass dwarf irregular galaxies (PGC\, 86669, 169957, 169968,
169969, and UGC\, 1561, 2905) with \hi line widths in the range
30--60~\kms\ FWHM. Three of these sources were in fact discovered in
the ``Arecibo Slice'' survey discussed below.
The seventh is a low luminosity spiral (UGC\,5672,
$M_B~=~-15.4$, M$_{HI}=5\times10^7$ M$_\odot$). None of these galaxies
appear to be fruitful starting points for a search for associated
low-mass satellite systems, since even the most massive is only about
0.01$L^*$. Since turn-around radius scales as the cube root of total
mass we expect turn-around radii of 55~kpc at most for these systems,
rather than the nominal value of 250~kpc for an $L^*$ galaxy. Exactly
one of the dwarf irregulars lies within 55~kpc of the AHISS track,
PGC\,169968.  At a distance of 8~Mpc, the AHISS 5$\sigma$ mass limit
is $2.2\times10^6$~M$_\odot$ over 32~\kms\ and the FWHM beam is some 7~kpc
broad. At an impact parameter of 17~kpc from an assumed 55~kpc radius
distribution, the strip samples a region of length 100~kpc. The
fractional area sampled by the AHISS coverage is then 18\%. The absence
of detected companions for this object in the AHISS implies that its
CHVC population should number less than about 10 for objects with mass
greater than $2.2\times10^6$~M$_\odot$.

\subsection{The ``Arecibo Slice'' Survey} 
\label{sec:aslice}

Another recent \hi survey is that of Spitzak \& Schneider
(\cite{spit98}) who utilized the Arecibo telescope to survey the region
between right ascension $22^{\rm h}$ to $03^{\rm h}24^{\rm m}$ and
$\delta=22^\circ58'$ to $\delta=23^\circ47'$. The flux sensitivity at
the hexagonally spaced sampling points was typically 1.7~\mjyb\ over
16~\kms\ at V$_{\rm Hel} = 5300$~\kms, decreasing to 81\% of this value
at 8340~\kms and to 56\% at 100~\kms. Since the sampling points were
spaced quite coarsely relative to the beam size (4.1 arcmin spacing and
a 3.3 arcmin beam), the flux sensitivity is estimated to vary by about
a factor of three, depending on whether sources were located at
sampling points or exactly between them. The corresponding average
5$\sigma$ mass sensitivity at a distance of 5~Mpc is about
$4\times10^6$~M$_\odot$, increasing to $1.5\times10^7$~M$_\odot$ at
10~Mpc. We searched for galaxy groups in the Garcia (\cite{garc93})
catalog in the velocity range +200 to +1200~\kms\ within a projected
seperation of 1~Mpc of the survey coverage, but found none. The only
four LEDA cataloged galaxies in this velocity range which lie inside
the survey coverage or within a projected turn-around radius are
PGC\,169957, 169968, 169969 and UGC\,1561. These are all very low mass
systems as already noted above in connection with the AHISS survey
discussion, where the same galaxies were also encountered.  The first
three of these objects were discovered in the ``Arecibo Slice'' survey.
Faint optical counterparts are seen for the PGC\,169968 and 169957,
which have \hi masses of about 1.3 and 3$\times10^7$~M$_\odot$
respectively. PGC\,169969 is the lowest \hi mass object found, in this
and any other blind \hi survey to date, of only about
$8\times10^6$~M$_\odot$. No optical counterpart to this source has yet
been detected, although a relatively bright fore-ground star makes this
a difficult undertaking. These objects fit into the more general trend
implied by Fig.~9 of Spitzak \& Schneider (\cite{spit98}) for \hi
selected objects to be increasingly gas-dominated at low galaxy mass.

\subsection{HIPASS survey of the Cen~A galaxy group}
\label{sec:hipass}

A portion of the Cen~A galaxy group has recently been subjected to a
similar search by Banks et al. (\cite{bank99}) using the HIPASS data
obtained with the Parkes 64~m telescope. They searched an area of 600
square degrees, corresponding to about 2.2~Mpc$^2$, for \hi detections
in the velocity range V$_{Hel}$~=~200 -- 1000~\kms. The 5$\sigma$ mass
sensitivity over 35~\kms\ was $7\times10^6$~M$_\odot$ for a nominal
distance of 3.5~Mpc, while the 15~arcmin angular resolution corresponds
to 15~kpc. This mass sensitivity has substantial overlap with the mass
distribution we derive for the Local Group CHVC population, although
a modest loss in total flux might be expected in the 15~kpc beam. The Cen~A
group has six major members within a total extent of about 1.9~Mpc,
making it about twice as rich as the Local Group. Banks et al. detect
10 new group members in their survey area, most of which have M$_{HI}$
near the $\sim10^7$~M$_\odot$ sensitivity limit. Five of these were not
previously cataloged, despite deep optical searches in this region, and
have extremely low optical luminosities (below M$_B \sim -15$) and
surface brightness ($<\mu_B> \sim 26$~mag arcsec$^2$). These five
objects have an average \hi mass of $2.0\pm0.8\times10^7$~M$_\odot$ and
a FWHM linewidth of 34$\pm$4~\kms. While this is still at the high mass
end of the Local Group CHVC distribution, it is possible that a new
population of low mass, extremely star-poor systems is starting to
emerge. An even more extreme object in this class seems to the possible
Cen~A group member HIPASS J1712$-$64 discovered by Kilborn et
al. (\cite{kilb00}). With V$_{Hel}$~=~460~\kms\ it is well-removed in
velocity from any confusing emission and in agreement with that of the
group. At an estimated distance of 3.2~Mpc, it has an \hi mass of
$1.7\times10^7$~M$_\odot$, 15~kpc size (at 10$^{19}$cm$^{-2}$ column
density) and lies at a projected seperation of 1.1~Mpc
from the massive Circinus galaxy. No optical counterpart to this
object has been detected down to a limiting surface brightness of
$\mu_B \sim 27$~mag arcsec$^2$.

\subsection{Targeted Survey of Groups} 
\label{sec:tsg}

A targeted survey of galaxy group environments was recently carried out
by Zwaan (\cite{zwaa00b}). A sparsely sampled grid of 60 pointings
was observed with the Arecibo telescope centered on each of five galaxy
groups at recession velocities of 1770, 1870, 2280, 2910 and 3000~\kms\
(corresponding to distances between about 27 and 46~Mpc). The 60
pointings were distributed over rectangular areas of about
2.5$\times$1.5~Mpc. At these distances each FWHM beam has an area of
24--40~kpc. Together, each grid of 60 pointings covered an area varying
from 0.027--0.075~Mpc$^2$, corresponding to a fractional coverage of
between 0.7--2.0\% for the various groups sampled. An average flux
sensitivity of 0.75~\mjyb\ over 10~\kms\ was obtained, corresponding to
5$\sigma$ mass limits 1.1--3.2$\times10^7$~M$_\odot$ over 30~\kms\ at
each beam center. Since the survey area is sparsely sampled we must
consider the sensitivity averaged over the beam area, rather than
simply taking the peak value.  The average 5$\sigma$ mass sensitivity
over each beam FWHM then varies between 1.6--4.6$\times10^7$~M$_\odot$
over 30~\kms. Only in the case of the nearest two galaxy groups is
there a small degree of overlap of the achieved mass sensitivity with
the expected CHVC mass distribution.  The constraint that follows from
non-detections over 0.75\% of the group area, is that there must be
less than about 100 objects with \hi mass exceeding
1.7$\times10^7$~M$_\odot$ associated with the NGC\,5798 and NGC\,5962
galaxy groups. Unfortunately this is not a very strong constraint.

\section{Comparison with other work}
\label{sec:comp}

Zwaan \& Briggs (\cite{zwaa00}) discuss detection statistics of
primordial gas clouds near galaxies and galaxy groups based on the
AHISS data. They state that the survey ``...  probed the haloes of
$\sim$300 cataloged galaxies and the environments of $\sim$14 groups
with sensitivity to neutral hydrogen masses $\ge10^7$~M$_\odot$.'' As
we have shown, a limiting \hi mass of $10^7$~M$_\odot$ is indeed a
minimal requirement for the expected detection of such features, and
even this will only allow detection of the most massive objects in the
expected distribution.  The AHISS survey attributes (Sorar
\cite{sora94}) lead to a distance limit of 15~Mpc for the 5$\sigma$
detection of an \hi mass of $10^7$~M$_\odot$ within the 15~kpc Arecibo
beam over the expected linewidth of 32~\kms. Out to this distance the
AHISS survey coverage passes within the turn-around radius of exactly
one galaxy group (the NGC\,628 group) and one cataloged galaxy
(PGC\,169968). Relaxing this distance limit to 20~Mpc, brings only one
additional galaxy group (the NGC\,3227 group) and no additional
galaxies into range.  

There are several reasons for this apparent discrepancy. The single
most important factor is the assumed distance, and hence mass, of the
population being sought. Assuming a baryon-to-dark-mass fraction,
$f$~=~0.1, and calculating a stability distance for the entire Wakker
\& van Woerden (\cite{wakk91c}) HVC sample following Blitz et
al. (\cite{blit99}) leads to a typical distance of 1~Mpc and \hi masses
in the range $10^7$--$10^{8.5}$~M$_\odot$. We have shown above that current
distance estimates to individual CHVCs in the Local Group lie in the
range 150--850~kpc and that the corresponding \hi mass range for CHVCs
cataloged in BB99 is $M_{\rm HI}=10^{5.7}\rightarrow10^{7.2}$
M$_\odot$.

Zwaan \& Briggs also assume a nominal 5$\sigma$ column density limit of
10$^{18}$cm$^{-2}$, corresponding to a flux density rms of 0.75~\mjyb
over 16~\kms\ that applies uniformly to a 180~arcsec wide strip, in
assessing the survey detectability of an extragalactic CHVC
population. The AHISS flux density rms is in fact typically higher than
0.75~\mjyb over 16~\kms\ as discussed specifically above, especially at
low velocities. In addition, the assumed linewidth of 16~\kms\ is about
a factor of two lower than the measured linewidths of this class of
object, making actual detection more difficult by a factor of
$\sqrt2$. Furthermore, the average detected response across the
180~arcsec survey strip is reduced by about 50\% for sources which are
comparable in size to the beam. Together these factors degrade the
effective detection sensitivity by more than a factor of three.

And finally, Zwaan \& Briggs have assumed that each cataloged field
galaxy, independant of mass, might host a comparably large population
of associated satellites extending out to a radius of 1~Mpc. In fact,
it is reasonable to expect (eg. Klypin et al. \cite{klyp99}) that the
number of associated low mass satellites might be proportional to the
host mass and that the radius over which they might extend be the
turn-around radius of the host mass (which scales as $M^{1/3}$).

\section{Conclusions}

Recent distance estimates for a sample of ten compact high-velocity
clouds (CHVCs) yield values in the range 150 to 850~kpc (Burton et
al. \cite{burt00}). The corresponding \hi masses of the BB99 catalog of
CHVCs range between $10^{5.7}\rightarrow10^{7.2}$ M$_\odot$ over an
average linewidth of 30~\kms. Interferometric surveys of isolated
fields have not generally achieved the combination of sensitivity and
sky coverage to put strong constraints on CHVC populations in galaxy
groups. The searches for low mass companions reported earlier by Taylor
et al. (\cite{tayl95}, \cite{tayl96}) and more recently by Pisano \&
Wilcots (\cite{pisa99}, \cite{pisa00}) fall short of probing the
expected CHVC mass range.

The most sensitive large--area \hi surveys are just beginning to sample
the upper mass end of the implied CHVC mass distribution.  The AHISS
survey (Zwaan et al. \cite{zwaa97}) could have detected objects with
$M_{\rm HI}=10^{7}$ M$_\odot$ out to a maximum distance of
15~Mpc. Unfortunately, out to this distance, the AHISS sky coverage
passes within the turn-around radius of only a single galaxy group and
a single isolated galaxy. The AHISS non-detections allow placing an
upper limit of about 300 objects exceeding $M_{\rm
HI}=5.7\times10^6$~M$_\odot$ within a radius of 1~Mpc of the NGC\,628
galaxy group. A targeted search of the environments of several galaxy
groups (Zwaan \cite{zwaa00b}) also falls short of reaching the required
mass sensitivity. The Arecibo Slice survey (Spitzak \& Schneider
\cite{spit98}) has detected several objects which overlap in \hi mass
with the CHVCs. The least massive of these ($M_{\rm
HI}\sim8\times10^{6}$ M$_\odot$) has no detected optical counterpart,
in keeping with a more general trend for low mass systems to be
increasingly gas-dominated.

The HIPASS survey has allowed detection of a small number of low mass
galaxies in the Cen~A galaxy group with extremely low surface
brightness optical counterparts ($<\mu_B> \sim 26$~mag
arcsec$^2$). This was accomplished with a 5$\sigma$ \hi mass
sensitivity of about $7\times10^6$~M$_\odot$ (Banks et
al. \cite{bank99}). In addition, at least one object has been found
with no detected optical counterpart down to $\mu_B \sim 27$~mag
arcsec$^2$ (Kilborn et al. (\cite{kilb00}). These objects appear to be
the highest mass counterparts of the Local Group CHVC population.

If the estimated distances of the Local Group CHVC population discussed
in \S\ref{sec:pops} are correct then pushing the detection limits down
by another order of magnitude, to $M_{\rm HI}=10^{6}$ M$_\odot$ (over
15~kpc and 30~\kms) will enable detection of populations comparable to
those determined for the Local Group, implying increased numbers by
some two orders of magnitude. If these low mass populations are not
found, then the Local Group hypothesis for the CHVCs must be seriously
reconsidered.

\begin{acknowledgements} 
  
We are grateful to Martin Zwaan for providing details of the
statistical analysis presented by Zwaan and Briggs (\cite{zwaa00}). We
acknowledge extensive use of the LEDA database, www-obs.univ-lyon1.fr.
This research has made use of the NASA/IPAC Extragalactic Database
(NED) which is operated by the Jet Propulsion Laboratory, California
Institute of Technology, under contract with the National Aeronautics
and Space Administration.

\end{acknowledgements}


\end{document}